\documentclass[amsmath,notitlepage,aps,prd,12pt,onecolumn]{revtex4-1}

\usepackage{amssymb}
\usepackage{amsmath}
\usepackage{multirow}
\usepackage{hyperref}
\usepackage{xspace}

\newcommand{\jpsi}{J/\psi}
\newcommand{\xicc}{\Xi_{cc}^+}
\newcommand{\Jpsi}{$J/\psi$\xspace}
\newcommand{\ccbar}{c\bar{c}}
\newcommand{\chic}[1][]{\chi_{c#1}}
\newcommand{\Chic}[1][]{$\chi_{c#1}$\xspace}
\newcommand{\AlpS}{$\alpha_S$\xspace}
\newcommand{\alpS}{\alpha_S}
\newcommand{\sigeff}{\sigma_\text{eff.}}
\newcommand{\pdf}{PDF}

\newcommand{\nb}{~\text{nb}}
\newcommand{\ub}{~\mathrm{\mu b}}
\newcommand{\mb}{~\text{mb}}

\newcommand{\GeV}{~\text{GeV}}
\newcommand{\TeV}{~\text{TeV}}

\begin{document}
\title{Production of quarkonia and doubly heavy baryons in $pp$-collisions with duality approach}
\author{V.V. Kiselev}
\email{Valery.Kiselev@ihep.ru}
\author{A.A. Novoselov}
\email{Aleksey.Novoselov@ihep.ru}
\author{E.R. Tagiev}
\email{Emin.Tagiev@ihep.ru}
\affiliation{Institute for High Energy Physics NRC ``Kurchatov Institute'', 142281, Protvino, Russia}
\affiliation{Moscow Institute of Physics and Technology, 141701, Dolgoprodny, Russia}

\begin{abstract}
In present work we discuss the production of heavy quarkonia and diquarks in $pp$-interactions.
The aim is to take into account the production of bound states of quarks originating from independent parton scatterings. 
This production is regulated by constraints on the invariant mass of constituents.
Such an approach leads to larger value of the diquark production cross section than 
traditional consideration in single parton scattering. This production mechanism of doubly heavy baryons can sooner be verified 
by modern experiments (e.g. LHCb). We also discuss related contributions to 
double and associated quarkonium production.
\end{abstract}
%\pacs{13.85.Fb, 14.40.R}

\maketitle

\section{Introduction}

The production of heavy quarks and their bound states in high energy hadron interactions is a perfect instrument for 
thorough studies of QCD. As heavy quark mass is substantially larger than the $\Lambda_{QCD}$ scale the 
production of heavy quark-antiquark pair is described by the perturbative QCD, while its hadronization allows 
rigorous theoretical description due to the non relativistic nature of considered bound states.

The constituent quark model predicts the existence of the baryon multiplet, which includes baryons with more than one heavy quark.
Currently only lightest of them $\xicc=ccd$ is claimed to be observed by the SELEX collaboration \cite{Mattson:2002vu,Ocherashvili:2004hi}.
However the cross section reported exceeds substantially the theoretical estimates~\cite{Kiselev:2001fw}. This result has not been 
confirmed by latter experimental searches~\cite{Ratti:2003ez,Aubert:2006qw,Chistov:2006zj}. 
Most recent experimental search in the $\xicc \to \Lambda_c^+ K^- \pi^+$
decay mode was performed by the LHCb collaboration \cite{Aaij:2013voa}.

Recent results on $c$-hadrons and quarkonium production unambiguously suggest that double parton scattering (DPS) contributes 
greatly to multiple production of hadrons with heavy flavours. One of the first experimental indications was obtained in 2012 in the
analysis of associated production of \Jpsi and open charm hadrons \cite{Aaij:2012dz}. The analysis of $\Upsilon$ and open charm 
\cite{Aaij:2015wpa} clearly claimed DPS to be the main source of signal. 

The natural question arises, whether DPS can contribute to the production of doubly heavy baryons. If yes, the hadronisation has to involve
heavy quarks produced in different parton subprocesses. It brings us to idea of the quark-hadron duality usage. Moreover, such an approach
has already been applied for the $e^+e^-$ case in \cite{Kiselev:1994pu}. However direct application of this approach leads to overestimation
of related cross sections (such as $pp\to \jpsi + \ccbar +X$). Thus we perform a calibration consisting in matching the calculations 
in duality approach and calculations which are guided by the quarkonia matrix elements (ME-based). The latter are known from the non-relativistic 
potential models (for all the considered states) and from the decay widths (for quarkonia states).

In the following section we consider associated production of \Jpsi and open charm. Third section is devoted to the production of diquarks.
Discussion of the results obtained is given in the last section. 
%We draw out conclusions in the last section. 

\section{\Jpsi plus open charm production}

Let us begin with calculation of $gg \to \jpsi + \ccbar$, $gg \to \psi(2S) + \ccbar$, and $gg \to \chic + \ccbar$ processes. 
In the leading order in \AlpS there are
$36$ Feynman diagrams for the $gg \to \ccbar \ccbar$ process. 

To generate amplitudes of these diagrams we use the 
FeynArts package~\cite{Mertig:1990an,Shtabovenko:2016sxi}. 
Let $p$ be the quarkonium momentum and $q$ --- the relative quark momentum, then
heavy quark and antiquark momenta can be denoted by
\begin{equation}
p_1 = p/2 + q ~\text{and}~
p_2 = p/2 - q,
\end{equation}
respectively. 
Following \citep{Bodwin:2002hg} we use the following spin projector operators for the considered spin-triplet charmonium states:
\begin{equation}
\Pi = \sum_{\lambda_1,\lambda_2} \frac{-1}{2\sqrt{2}(E+m)}\bar{v}(p_2,\lambda_2) \epsilon^\lambda_\mu \gamma^\mu \frac{\hat{p}+2E}{2E}u(p_1,\lambda_1),
\end{equation}
where $m$ and $E$ are quark mass and energy, $\lambda_{1,2}$ are quark helicities. For the $S$-wave 
\Jpsi and $\psi(2S)$ states $p_1=p_2$ and the spin projector operator simplifies as follows:
\begin{equation}
\Pi_S = \frac{-1}{2\sqrt{2}} (\hat{p} - 2 m) \hat{\epsilon},
\label{Eqn:SwaveProj}
\end{equation}
where we also substituted $m$ for $E$ due to the non relativistic treatment. Here $\epsilon$ is the spin polarization vector
of the quarkonium, $\epsilon\cdot\epsilon^*=-1$ and $\epsilon\cdot p=0$. For the $P$-wave 
\Chic states we have:
\begin{equation}
\Pi_P = \frac{-1}{8\sqrt{2}~m^2} (\hat{p}/2 - \hat{q} - m) (\hat{p} - 2 m) \hat{\epsilon} (\hat{p}/2 + \hat{q} + m),
\end{equation}

Concerning the color part of projectors, we consider only color singlet (CS) production which is expressed by 
$\delta_{ij}/\sqrt{N_c} = \delta_{ij}/\sqrt{3}$
color projector operator.

For the numerical calculation of hadronic cross section we use CT14llo \pdf s at the factorization scale 
$\mu = \sqrt{M^2+2m^2+p_T^2}$, where the quarkonium mass $M$ is taken equal $2m$. 
Leading order expression for the strong coupling $\alpha_S(\mu)$ is 
used at the same scale. Numerical values of these quantities are obtained from the LHAPDF package~\cite{Buckley:2014ana}.

We use $m=m_c=1.5\GeV$ as the heavy quark mass and the following values of SC long distance matrix elements:
\begin{eqnarray}
\langle \mathcal{O}(1^3S_1^{[1]}) \rangle &=& \frac{|R_{\jpsi}(0)|^2}{4\pi} = 0.0447 \GeV^3,
\nonumber \\
\langle \mathcal{O}(2^3S_1^{[1]}) \rangle &=& \frac{|R_{\psi(2S)}(0)|^2}{4\pi} = 0.0269 \GeV^3,
\\
\langle \mathcal{O}(1^3P_J^{[1]}) \rangle &=& (2J+1)\frac{3 |R'_{\chic}(0)|^2}{4\pi} = (2J+1)\cdot 0.0179 \GeV^3,
\nonumber 
\end{eqnarray}
Here $|R(0)|^2$ and $|R'(0)|^2$ values are obtained from known decay widths of quarkonia.

For the $P$-wave states production we use HELAC-onia~\cite{Shao:2012iz,Shao:2015vga} at the parton level. 
Then we weight the produced events with our \pdf s and $\alpha_S(\mu)$ to get the hadronic cross sections. 
As HELAC-onia does not support LHAPDF6 there is no way to use it 
with the CT14 \pdf s directly. We have checked that results for $gg \to \jpsi + \ccbar$, $gg \to \jpsi + \jpsi$ 
obtained with the HELAC-onia and our code are in agreement. For the feed-down calculation we use branching 
fractions from~\cite{Olive:2016xmw}.
The resultant cross sections are presented in 
Table~\ref{Tab:CS}.

\begin{table}
\begin{tabular}{|c|c|c|c|c|}
\hline
\multirow{2}{*}{Final state}		& \multicolumn{2}{c|}{$\sqrt{s}_{pp} = 7 \TeV$} & \multicolumn{2}{c|}{$\sqrt{s}_{pp} = 13 \TeV$}	\\
\cline{2-5}
													&	~No cuts~	& $2<y<4.5$	& ~No cuts~		& $2<y<4.5$		\\
\hline
$\jpsi + \jpsi$, direct					& $13.4 \nb$	& $2.1 \nb$		& $23.4 \nb$	& $3.7 \nb$			\\
\hline
$\jpsi + \ccbar$, direct				& $0.55 \ub$	& $74 \nb$		& $0.95 \ub$	& $0.13 \ub$		\\
\hline
$\psi(2S) + \ccbar$					& $0.33 \ub$	& $45 \nb$		& $0.57 \ub$	& $79 \nb$			\\
\hline
$\chic[0] + \ccbar$					& $0.18 \ub$	& $24 \nb$		& $0.31 \ub$	& $44 \nb$			\\
\hline
$\chic[1] + \ccbar$					& $0.19 \ub$	& $26 \nb$		& $0.33 \ub$	& $46 \nb$			\\
\hline
$\chic[2] + \ccbar$					& $0.36 \ub$	& $49 \nb$		& $0.63 \ub$	& $87 \nb$			\\
\hline
$\jpsi + \ccbar$ with feed-down	& $0.87 \ub$	& $0.12 \ub$	& $1.53 \ub$	& $0.21 \ub$	\\
\hline
\end{tabular}
\caption{Hadronic cross sections for the subprocesses with charmonia production. ME-based calculation.
The rapidity restriction is applied to the charmonium and one of the $c$-quarks.
\label{Tab:CS}}
\end{table}

Next we calculate the charmonium production in the duality approach. The partonic subprocess is 
$gg\to \ccbar \ccbar$ again, but we do not apply any spin projectors. Only CS $\ccbar$ pairs are selected.
We take $m_c=1.4\GeV$~\cite{Kiselev:2001fw,Penarrocha:2001ig} and the duality region
$2m_c < M_{inv.} < 2m_D+\Delta$ with $m_D=1.86 \GeV$ and $\Delta = 0.5 \GeV$~\cite{Kiselev:2001fw}. 
Here $m_c$ differs from the ME-based calculation above to be consistent with other duality-based calculations.
The cross sections are summarized in Table~\ref{Tab:CSdual}. 

\begin{table}
\begin{tabular}{|c|c|c|}
\hline
Final state												& $\sqrt{s}_{pp} = 7 \TeV$	& $\sqrt{s}_{pp} = 13 \TeV$ 	\\
\hline
$\ccbar \ccbar$										& $21 \ub$								& $36 \ub$									\\
\hline
$(\ccbar)_{dual}^{[1]} + c + \bar{c}$	& $2.1 \ub$								& $3.7 \ub$									\\
\hline
\end{tabular}
\caption{Hadronic cross sections for the duality-based approach with the SPS source ($gg\to\ccbar\ccbar$ subprocess).
$(\ccbar)_{dual}^{[1]}$ is in the CS color state and has $M_{inv}$ in the duality region.
\label{Tab:CSdual}}
\end{table}

Comparison with the cross sections from Table~\ref{Tab:CS} leads to the conclusion that approximately 41\% of the CS quark-antiquark
pairs in the duality region transit into \Jpsi. If we account only direct \Jpsi production this fraction is about 26\%. 
%We consider the cross sections
%without cuts as with LHAb cuts we also force $c$- or $\bar{c}$-quark to be in the rapidity window for the $\jpsi + \ccbar$ case.

Let us now turn to the DPS production source. We have two $g g \to \ccbar$ partonic subprocesses. According to the phenomenological
expression for the DPS cross section,
\begin{equation}
\sigma(pp \to 2 \times \ccbar+X)=\frac{1}{2}\frac{\sigma(pp \to \ccbar+X)^2}{\sigeff}.
\label{Eqn:dpsDiCCbar}
\end{equation}
The most recent measurement of $\sigeff$ value was performed by LHCb in analysis of the $\Upsilon$ plus open charm 
production~\cite{Aaij:2015wpa}, $\sigeff=18\mb$. This leads to the following $pp\to2\times\ccbar+X$ cross sections
\begin{eqnarray}
\sigma(pp \to 2 \times \ccbar+X, 7 \TeV)&=&0.51\mb,
\nonumber \\
\sigma(pp \to 2 \times \ccbar+X, 13 \TeV)&=&1.2\mb,
\label{Eqn:CCCCdpsCS}
\end{eqnarray}
if the $g g \to \ccbar$ subprocesses is calculated in LO with $m_c = 1.4 \GeV$. We would like to notice that these DPS cross sections
grow more rapidly with the $\sqrt{s_{pp}}$ increase. It is the consequence of the assumption that $\sigeff$ depends weakly on the 
$\sqrt{s_{pp}}$.

From the generated DPS events we take $c$ and $\bar{c}$ quarks from the different $\ccbar$ pairs. We select events with $\ccbar$
invariant mass in the duality region and apply additional $1/9$ factor to the cross section to account only CS production. 
We also apply $0.41$ suppression factor which we earlier found to correspond the $J/\psi$ production with feed-down from
the other charmonium states. The resultant cross sections are presented in Table~\ref{Tab:CSdualDPS}. 

\begin{table}
\begin{tabular}{|c|c|c|c|c|}
\hline
\multirow{2}{*}{Final state} & \multicolumn{2}{c|}{$\sqrt{s}_{pp} = 7 \TeV$} & \multicolumn{2}{c|}{$\sqrt{s}_{pp} = 13 \TeV$}	\\
\cline{2-5}
							& ~No cuts~		& $2<y<4.5$	& ~No cuts~			& $2<y<4.5$	\\
\hline
$\ccbar \ccbar$	& $0.51 \mb$	&						& $1.2 \mb$		&						\\
\hline
$\jpsi + \ccbar$	& $6.4 \ub$		& $0.9 \ub$		& $13.8 \ub$		& $2 \ub$		\\
\hline
$\jpsi + \jpsi$		& $62 \nb$		& $7 \nb$		& $0.13 \ub$		& $15 \nb$		\\
\hline
\end{tabular}
\caption{Hadronic cross sections for the duality-based approach with the DPS source of $\ccbar$-pairs. 
The rapidity restriction is applied to the charmonium and one of the $c$-quarks.
\label{Tab:CSdualDPS}}
\end{table}

%(*I have no thoughts why Di-psi in LHCb is bigger then measured...*)

The cross section of the $pp\to \jpsi \jpsi + X $ process obtained with our DPS plus duality approach is close to those measured by LHCb.
However there should be contributions from the $gg \to \jpsi \jpsi$ subprocess and from the ordinary DPS. First one is presented 
in Table~\ref{Tab:CS}. The latter is known from the measured prompt \Jpsi production. Currently the sum of these three contributions 
exceeds the measured value. Let us discuss possible reasons more in details.

The ordinal DPS cross section for double \Jpsi is represented by a simple expression:
\begin{equation}
\sigma(pp \to 2 \times \jpsi + X)=\frac{1}{2}\frac{\sigma(pp \to \jpsi + X)^2}{\sigeff}.
\end{equation}
As in the expression for $\sigma(pp \to 2 \times \ccbar + X)$~(\ref{Eqn:dpsDiCCbar}) we take $\sigeff$ measured by LHCb in
the associated $\Upsilon$ and open charm production~\cite{Aaij:2015wpa}, $\sigeff=18\mb$. The extraction of $\sigeff$ relies 
on the assumption that all signal originates from the DPS contribution. The measured kinematic distributions support this assumption. 
This value of $\sigeff$ exceeds measured in other processes at same $\sqrt{s_{pp}}$. It also supports the assumption that no
other sources contribute significantly to the final state considered. 
%What concerns the duality production from DPS it can not
%contribute to the $\Upsilon$ plus open charm ...
Taking measured cross sections of prompt \Jpsi 
production~\cite{Aaij:2011jh,%Aaij:2013yaa,
Aaij:2015rla} one gets the cross sections for double \Jpsi presented in
the first row of Table~\ref{Tab:CSdiPsi}.

\begin{table}
\begin{tabular}{|c|c|c|}
\hline
Source 									& $\sqrt{s}_{pp} = 7 \TeV$	& $\sqrt{s}_{pp} = 13 \TeV$ 	\\
\hline
Ordinal DPS							& $3.1 \nb$								& $6.5 \nb$									\\
\hline
SPS ($gg\to\jpsi\jpsi$)		& $2.1 \nb$								& $3.7 \nb$									\\
\hline
SPS (with feed-down)			& $2.9 \nb$								& $5.1 \nb$									\\
\hline
DPS+duality							& $7 \nb$								& $15 \nb$									\\
\hline
\hline
Sum										& $13.0 \nb$							& $21.6 \nb$								\\
\hline
\hline
Measured								& $5.1 \nb$								& in progress								\\
\hline
\end{tabular}
\caption{Dual \Jpsi production in the LHCb fiducial region.
\label{Tab:CSdiPsi}}
\end{table}

What concerns the SPS contribution, we considered the $gg \to \jpsi \jpsi$ subprocess at the LO in \AlpS. There can be feed-down
from the $\psi(2S)$ and \Chic states. Feed-down from the $\psi(2S)$ was considered in our earlier 
works~\cite{Berezhnoy:2011xy,Berezhnoy:2012xq} and is about $1/3$ (see Table~\ref{Tab:CSdiPsi}). 
Feed-down from $\jpsi+\chic$ final state was also considered. In~\cite{Likhoded:2016zmk} it was shown that it is surprisingly small 
and has the order of picobarns. The LO $gg \to \jpsi \jpsi$ cross section in~\cite{Likhoded:2016zmk} differs significantly from
those in Table~\ref{Tab:CSdiPsi} because of the use of NLO CT10 \pdf s for both NLO and LO calculations. Corrections due to 
the real gluon emission were also studied in~\cite{Lansberg:2014swa}. We adhere to an opinion that the LO contributions dominate 
for the total cross section while \AlpS{}-corrections are crucial at high $p_T$. In earlier works we used CTEQ5L and CTEQ6L1 LO \pdf s.
Currently we switched to the most modern CTEQ+TEA LO set -- CT14llo. These \pdf\ updates gradually decreased the cross section 
prediction. The fiducial region of the LHCb detector corresponds to the $y \in [2,4.5]$ region. The $x=x_1x_2$ value is of order $10^{-6}$.
Thus we are sensitive to the \pdf\ values at $x_1 \approx 10^{-4}$. The refining of \pdf s from CTEQ5L to CT14llo leads to factor of $2$ 
decrease in SPS $pp\to 2\times \jpsi + X$ cross section prediction. There are studies of gluon saturation effects in this $x$ 
region~\cite{Watanabe:2015yca}, which suggest possible further cross section decrease. 

The contribution of the duality approach production from the DPS source depends on the maximum number of assumptions. 
It is again sensitive 
to the $\sigeff$ parameter of DPS estimation. More uncertainty comes from the selection of duality region including selection 
of the $m_c$. We ensure that selected duality region together with the suppression factor, which corresponds to the \Jpsi 
production, reproduce the cross section of $gg \to \jpsi \ccbar$ partonic subprocess. However there can be influence 
from different distribution over the $(\ccbar)$ invariant mass in DPS and in $gg \to \ccbar \ccbar$ subprocess. The argument
in favour of such an approach is that \Jpsi production in this case is correlated with the production of open charm hadrons
even with DPS source. The experimental measurement~\cite{Aaij:2012dz} of associated production of \Jpsi and 
open charm hadrons indeed demonstrates that the $p_T$ distribution of \Jpsi differs from the prompt \Jpsi distribution.
It can be the sequence of the kinematic cut on the minimum $p_T$ of the charm hadron produced and the correlated influence
on the \Jpsi $p_T$-spectrum. It is not the case for the $\Upsilon$ and open charm associated production~\cite{Aaij:2015wpa}. 
Cross sections in both measurements can only be interpreted if DPS source is involved.

The actual problems and results on double \Jpsi production have been recently addressed in~\cite{Lansberg:2016chs}.

\section{($cc$)-diquark production}

As we mentioned in the introduction the ($cc$)-diquark production studies are necessary to estimate yield of double 
heavy baryons in the LHC conditions. The customary consideration~\citep{Kiselev:2001fw} identify heavy
diquark as a static color source which is surrounded by the light quark. Indeed as the mass of the charm quark is 
substantially larger than the $\Lambda_{QCD}$ scale, two charm quarks form a compact 
diquark system, which is considerably smaller than the radius of the light quark confinement.

Contrary to the quarkonium production diquark formation should be followed by hadronization to obtain an 
observable state. The recombination and fragmentation scenarios can occur. It was shown that consideration 
of fragmentation diagrams only is not sufficient and recombination should dominate at low $p_T$~\citep{Kiselev:2001fw}. 
Currently there is no known way to account the probability for diquark to dissolve, which can be larger than for the 
quarkonium states due to the non-CS configuration. Thus the diquark production cross sections 
estimations presented in current section are the upper estimates for the doubly heavy baryon.

We consider production of the lightest doubly heavy baryon $\xicc$. As it has two identical $c$-quarks Pauli principle
restricts the diquark state to spin-triplet. For the spin projector operator we use same expression as for the $S$-wave 
quarkonia~(\ref{Eqn:SwaveProj}). The color projector of the $\bar{3}_c$ diquark state is $\epsilon_{ijk}/\sqrt{2}$
with $i, j$ being color indices of quarks, $k$ --- color index of diquark.

As for the $gg \to \jpsi \ccbar$ subprocess we start with amplitudes of the $gg \to \ccbar \ccbar$ diagrams. 
We use $m=m_c=1.5\GeV$, LO running \AlpS and CT14llo \pdf s at the scale $\mu = \sqrt{M^2+2m^2+p_T^2}$.
Mass $M$ of the $1S$ diquark is $3.16\GeV$~\cite{Kiselev:2001fw} and is equal to the $c$-quark mass doubled
at the same level of accuracy as for the \Jpsi, $\psi(2S)$ charmonia states.
The radial wave function value in origin is taken equal $R(0)=0.53\GeV^{3/2}$~\cite{Kiselev:2001fw}, so
\begin{equation}
\langle \mathcal{O}(1^3S_1^{[\bar{3}]}) \rangle = 0.022 \GeV^3.
\end{equation}
%For the $3$-particle phase space generation the PHEGAS~\cite{Papadopoulos:2000tt} program is used.
The hadronic cross section results are summarized in the first row of Table~\ref{Tab:CSdiq}.

\begin{table}
\begin{tabular}{|c|c|c|c|c|}
\hline
\multirow{2}{*}{Final state} 									& \multicolumn{2}{c|}{7 TeV} & \multicolumn{2}{c|}{13 TeV} 	\\
\cline{2-5}
																				& ~No cuts~		& $2<y<4.5$ 	& ~No cuts~		& $2<y<4.5$ 		\\
\hline
$(cc)_{1S}^{[\bar{3}]} + \bar{c} + \bar{c}$			& $0.4 \ub$		& $0.06 \ub$	& $0.7 \ub$		& $0.1 \ub$			\\
\hline
$(cc)_{dual}^{[\bar{3}]} + \bar{c} + \bar{c}$		& $1.5 \ub$		& 	$0.2 \ub$	& $2.6 \ub$		&	$0.4 \ub$		\\
\hline
\end{tabular}
\caption{Hadronic cross sections for the $(cc)^{[\bar{3}]}$ production. 
$(cc)_{dual}^{[\bar{3}]}$ is in the antitriplet color state and has $M_{inv}$ in the duality region.
The rapidity restriction is applied to the diquark only.
\label{Tab:CSdiq}}
\end{table}

Next we calculate the $(cc)^{[\bar{3}]}$ production in the duality approach. Subprocess $gg\to \ccbar \ccbar$ 
with only $\bar{3}_c$ color projector operators is considered. The $1/2$ factor in amplitude is inserted to take
the identity of quarks or antiquarks into account. The $(cc)^{[\bar{3}]}$-state invariant mass is restricted
to the duality region. As for quarkonia we take $m_c=1.4\GeV$ and the duality region
$2m_c < M_{inv.} < 2m_D+\Delta$ with $m_D=1.86 \GeV$ and $\Delta = 0.5 \GeV$~\cite{Kiselev:1994pu}. 
The cross sections are presented in the second row of Table~\ref{Tab:CSdiq}. The fraction of $\bar{3}_c$-states 
in the duality region that correspond to the $^{1S}(cc)_{\bar{3}}$ is found equal $0.26$.

Finally we turn to the DPS source of $c$ or $\bar{c}$-pairs. The cross sections for this source of considered final state
were written down in~(\ref{Eqn:CCCCdpsCS}). From the generated DPS events we select those with invariant mass 
of $(cc)$ or $(\bar{c}\bar{c})$ being in the duality range. Additional $1/3$ factor accounts formation of the
antitriplet states only. We also apply suppression by factor of $0.26$ found above. 
The resultant cross sections are presented in Table~\ref{Tab:CSdiqDPS}. They are by approximately order of 
magnitude larger than in the ME-based approach.

\begin{table}
\begin{tabular}{|c|c|c|c|c|}
\hline
\multirow{2}{*}{Final state} 							& \multicolumn{2}{c|}{7 TeV}	& \multicolumn{2}{c|}{13 TeV}	\\
\cline{2-5}
																		& ~No cuts~		& $2<y<4.5$ 	& ~No cuts~			& $2<y<4.5$ 	\\
\hline
$(cc)_{1S}^{[\bar{3}]} + \bar{c} + \bar{c}$	& $3 \ub$		& $0.5 \ub$		& $6.6 \ub$			& $1 \ub$		\\
\hline
\end{tabular}
\caption{Hadronic cross sections for the duality-based approach with the DPS source of $(cc)$-pairs. 
The rapidity restriction is applied to the diquark only.
\label{Tab:CSdiqDPS}}
\end{table}

\section{Discussion and conclusion}

This report is devoted to phenomenological analysis of the doubly heavy baryon production in $pp$-interaction. 
As it is stated in~\cite{Berezhnoy:1998aa}, this process is closely connected to the associated \Jpsi and
open charm production. If SPS processes were the dominant source of the $\jpsi+\ccbar$ signal the comparison with experiment
could eliminate uncertainties connected with $\alpS(\mu)$, \pdf s and $m_c$ selection. However recent
observations claim that DPS processes dominate in the production of many final states, which include several heavy
hadrons. This is also the case for the $\jpsi+\ccbar$ production. Currently the SPS and DPS contributions to
this final state production can not be separated. The work is more active for the $\jpsi+\jpsi$ case, where 
kinematic correlations provide better basis for separation.

Thus a dramatic difference between $\jpsi+\ccbar$ and $(cc)+\bar{c}\bar{c}$ production processes arises.
Due to the quark composition only first final state cross section can be estimated with the customary DPS treatment.
Succeeding with the interpretation of measured cross section, this treatment fails to account for the difference 
in $p_T$-spectra of associated and prompt \Jpsi-mesons. Interesting to mention that there is no such 
discrepancy in production of $\Upsilon$ plus open charm. We suppose that the reason for the discrepancy in the $\jpsi+\ccbar$ 
case is the existence of kinematic cut on the transverse momentum of the open charm hadron. The $p_T$-spectrum 
of \Jpsi is affected by it if there are correlations between production of the \Jpsi-meson and the accompaning 
$\ccbar$-pair. Such an interinfluence can arise 
from correlations between momenta of initial partons, which are not taken into account in the simple DPS
model. In such a case it should be observable through the alteration of the $p_T$-spectrum of $\Upsilon$-mesons with 
application of
more rough $p_T$-cut on the open charm hadrons in $\Upsilon$ and $\ccbar$ associated production. Another possibility is the 
involvement of quarks produced in different partonic subprocesses to the quarkonia formation. The  
$\Upsilon + \ccbar$ production is not influenced in this scenario. 

In our analysis we consider second scenario more in details. Apart from the influence on the $\jpsi+\ccbar$ production
features it can give rise to the doubly heavy baryon cross section. For the bound state formation from
quarks originating from different parton subprocesses one can not control the quantum numbers of quark pairs.
We suggest using the duality approach for this case. Apart from requirement for the invariant mass to be in 
the duality region one needs to apply some suppression factors to the cross section to correspond with the
formation of specific states. The correspondence is achieved by matching results obtained in the duality approach 
with those obtained in the standard ME-based formalism. It can be done in SPS production case as in it
both approaches describe the singe phenomenon. Then we apply duality-based calculation to the 
DPS source. The resultant cross section predictions for diquark production are about order of magnitude larger 
than those obtained in the ME-based calculation. 

All calculations of heavy diquark production cross section are to be considered as upper estimates for the 
doubly heavy baryon production. The interaction between diquark and gluons is not suppressed 
in contrary to the CS $\ccbar$ pairs, where the quarkonium dissociation supposes an exchange 
by two hard gluons with the quark-gluon sea. The gluon virtuality is to be of same order or greater than 
the inverse size of quarkonium.
In principle one can imagine that study of this phenomenon is possible by measuring suppression of 
doubly heavy baryon yield in the low-$p_T$ range with respect to the quarkonium plus open flavour yield but in 
practice the uncertainties involved ruin this opportunity. Indeed better understanding of parton shover in proton is
needed to obtain more rigorous predictions in the DPS formalism. 

We discussed uncertainties brought by the duality usage by the example of \Jpsi pair production.
Apart from the uncertainties connected with $\alpS(\mu)$, \pdf s and $m_c$ values one faces with the dependence on 
the duality region selection. Even with introduction of the suppression factor which provides matching with the ME-based calculation 
there is dependence on the invariant mass distribution shape, which is different for SPS and DPS production sources.

Despite mentioned difficulties we consider the mechanism discussed as quite feasible. DPS is also known 
to give increase to the cross sections of many processes with multiple heavy quark production. The cross sections
predicted give hope the $\xicc$-baryons will soon be observed by the LHC experiments. Additional insights will 
be provided by updated measurements of double and associated quarkonia production.

\acknowledgements
The authors would like to thank Prof. Anatoly Likhoded and Dr. Stanislav Poslavsky for fruitful discussions. 
The work of A.A.N. and E.R.T. was supported by the Grant MK-6645.2015.2. 
The work of V.V.K. and A.A.N. was supported by the Russian Foundation of Basic Research grant \#15-02-03244.

\bibliographystyle{apsrev4-1}
\bibliography{main}

\begin{thebibliography}{27}%
\makeatletter
\providecommand \@ifxundefined [1]{%
 \@ifx{#1\undefined}
}%
\providecommand \@ifnum [1]{%
 \ifnum #1\expandafter \@firstoftwo
 \else \expandafter \@secondoftwo
 \fi
}%
\providecommand \@ifx [1]{%
 \ifx #1\expandafter \@firstoftwo
 \else \expandafter \@secondoftwo
 \fi
}%
\providecommand \natexlab [1]{#1}%
\providecommand \enquote  [1]{``#1''}%
\providecommand \bibnamefont  [1]{#1}%
\providecommand \bibfnamefont [1]{#1}%
\providecommand \citenamefont [1]{#1}%
\providecommand \href@noop [0]{\@secondoftwo}%
\providecommand \href [0]{\begingroup \@sanitize@url \@href}%
\providecommand \@href[1]{\@@startlink{#1}\@@href}%
\providecommand \@@href[1]{\endgroup#1\@@endlink}%
\providecommand \@sanitize@url [0]{\catcode `\\12\catcode `\$12\catcode
  `\&12\catcode `\#12\catcode `\^12\catcode `\_12\catcode `\%12\relax}%
\providecommand \@@startlink[1]{}%
\providecommand \@@endlink[0]{}%
\providecommand \url  [0]{\begingroup\@sanitize@url \@url }%
\providecommand \@url [1]{\endgroup\@href {#1}{\urlprefix }}%
\providecommand \urlprefix  [0]{URL }%
\providecommand \Eprint [0]{\href }%
\providecommand \doibase [0]{http://dx.doi.org/}%
\providecommand \selectlanguage [0]{\@gobble}%
\providecommand \bibinfo  [0]{\@secondoftwo}%
\providecommand \bibfield  [0]{\@secondoftwo}%
\providecommand \translation [1]{[#1]}%
\providecommand \BibitemOpen [0]{}%
\providecommand \bibitemStop [0]{}%
\providecommand \bibitemNoStop [0]{.\EOS\space}%
\providecommand \EOS [0]{\spacefactor3000\relax}%
\providecommand \BibitemShut  [1]{\csname bibitem#1\endcsname}%
\let\auto@bib@innerbib\@empty
%</preamble>
\bibitem [{\citenamefont {Mattson}\ \emph {et~al.}(2002)\citenamefont {Mattson}
  \emph {et~al.}}]{Mattson:2002vu}%
  \BibitemOpen
  \bibfield  {author} {\bibinfo {author} {\bibfnamefont {M.}~\bibnamefont
  {Mattson}} \emph {et~al.} (\bibinfo {collaboration} {SELEX}),\ }\href
  {\doibase 10.1103/PhysRevLett.89.112001} {\bibfield  {journal} {\bibinfo
  {journal} {Phys. Rev. Lett.}\ }\textbf {\bibinfo {volume} {89}},\ \bibinfo
  {pages} {112001} (\bibinfo {year} {2002})},\ \Eprint
  {http://arxiv.org/abs/hep-ex/0208014} {arXiv:hep-ex/0208014 [hep-ex]}
  \BibitemShut {NoStop}%
%%CITATION = HEP-EX/0208014;%%
\bibitem [{\citenamefont {Ocherashvili}\ \emph {et~al.}(2005)\citenamefont
  {Ocherashvili} \emph {et~al.}}]{Ocherashvili:2004hi}%
  \BibitemOpen
  \bibfield  {author} {\bibinfo {author} {\bibfnamefont {A.}~\bibnamefont
  {Ocherashvili}} \emph {et~al.} (\bibinfo {collaboration} {SELEX}),\ }\href
  {\doibase 10.1016/j.physletb.2005.09.043} {\bibfield  {journal} {\bibinfo
  {journal} {Phys. Lett.}\ }\textbf {\bibinfo {volume} {B628}},\ \bibinfo
  {pages} {18} (\bibinfo {year} {2005})},\ \Eprint
  {http://arxiv.org/abs/hep-ex/0406033} {arXiv:hep-ex/0406033 [hep-ex]}
  \BibitemShut {NoStop}%
%%CITATION = HEP-EX/0406033;%%
\bibitem [{\citenamefont {Kiselev}\ and\ \citenamefont
  {Likhoded}(2002)}]{Kiselev:2001fw}%
  \BibitemOpen
  \bibfield  {author} {\bibinfo {author} {\bibfnamefont {V.~V.}\ \bibnamefont
  {Kiselev}}\ and\ \bibinfo {author} {\bibfnamefont {A.~K.}\ \bibnamefont
  {Likhoded}},\ }\href {\doibase 10.1070/PU2002v045n05ABEH000958} {\bibfield
  {journal} {\bibinfo  {journal} {Phys. Usp.}\ }\textbf {\bibinfo {volume}
  {45}},\ \bibinfo {pages} {455} (\bibinfo {year} {2002})},\ \bibinfo {note}
  {[Usp. Fiz. Nauk172,497(2002)]},\ \Eprint
  {http://arxiv.org/abs/hep-ph/0103169} {arXiv:hep-ph/0103169 [hep-ph]}
  \BibitemShut {NoStop}%
%%CITATION = HEP-PH/0103169;%%
\bibitem [{\citenamefont {Ratti}(2003)}]{Ratti:2003ez}%
  \BibitemOpen
  \bibfield  {author} {\bibinfo {author} {\bibfnamefont {S.~P.}\ \bibnamefont
  {Ratti}},\ }\bibfield  {booktitle} {\emph {\bibinfo {booktitle}
  {{Proceedings, 5th International Conference on Hyperons, charm and beauty
  hadrons (BEACH 2002): Vancouver, Canada, June 25-29, 2002}}},\ }\href
  {\doibase 10.1016/S0920-5632(02)01948-5} {\bibfield  {journal} {\bibinfo
  {journal} {Nucl. Phys. Proc. Suppl.}\ }\textbf {\bibinfo {volume} {115}},\
  \bibinfo {pages} {33} (\bibinfo {year} {2003})},\ \bibinfo {note}
  {[,33(2003)]}\BibitemShut {NoStop}%
%%CITATION = NUPHZ,115,33;%%
\bibitem [{\citenamefont {Aubert}\ \emph {et~al.}(2006)\citenamefont {Aubert}
  \emph {et~al.}}]{Aubert:2006qw}%
  \BibitemOpen
  \bibfield  {author} {\bibinfo {author} {\bibfnamefont {B.}~\bibnamefont
  {Aubert}} \emph {et~al.} (\bibinfo {collaboration} {BaBar}),\ }\href
  {\doibase 10.1103/PhysRevD.74.011103} {\bibfield  {journal} {\bibinfo
  {journal} {Phys. Rev.}\ }\textbf {\bibinfo {volume} {D74}},\ \bibinfo {pages}
  {011103} (\bibinfo {year} {2006})},\ \Eprint
  {http://arxiv.org/abs/hep-ex/0605075} {arXiv:hep-ex/0605075 [hep-ex]}
  \BibitemShut {NoStop}%
%%CITATION = HEP-EX/0605075;%%
\bibitem [{\citenamefont {Chistov}\ \emph {et~al.}(2006)\citenamefont {Chistov}
  \emph {et~al.}}]{Chistov:2006zj}%
  \BibitemOpen
  \bibfield  {author} {\bibinfo {author} {\bibfnamefont {R.}~\bibnamefont
  {Chistov}} \emph {et~al.} (\bibinfo {collaboration} {Belle}),\ }\href
  {\doibase 10.1103/PhysRevLett.97.162001} {\bibfield  {journal} {\bibinfo
  {journal} {Phys. Rev. Lett.}\ }\textbf {\bibinfo {volume} {97}},\ \bibinfo
  {pages} {162001} (\bibinfo {year} {2006})},\ \Eprint
  {http://arxiv.org/abs/hep-ex/0606051} {arXiv:hep-ex/0606051 [hep-ex]}
  \BibitemShut {NoStop}%
%%CITATION = HEP-EX/0606051;%%
\bibitem [{\citenamefont {Aaij}\ \emph {et~al.}(2013)\citenamefont {Aaij} \emph
  {et~al.}}]{Aaij:2013voa}%
  \BibitemOpen
  \bibfield  {author} {\bibinfo {author} {\bibfnamefont {R.}~\bibnamefont
  {Aaij}} \emph {et~al.} (\bibinfo {collaboration} {LHCb}),\ }\href {\doibase
  10.1007/JHEP12(2013)090} {\bibfield  {journal} {\bibinfo  {journal} {JHEP}\
  }\textbf {\bibinfo {volume} {12}},\ \bibinfo {pages} {090} (\bibinfo {year}
  {2013})},\ \Eprint {http://arxiv.org/abs/1310.2538} {arXiv:1310.2538
  [hep-ex]} \BibitemShut {NoStop}%
%%CITATION = ARXIV:1310.2538;%%
\bibitem [{\citenamefont {Aaij}\ \emph {et~al.}(2012)\citenamefont {Aaij} \emph
  {et~al.}}]{Aaij:2012dz}%
  \BibitemOpen
  \bibfield  {author} {\bibinfo {author} {\bibfnamefont {R.}~\bibnamefont
  {Aaij}} \emph {et~al.} (\bibinfo {collaboration} {LHCb}),\ }\href {\doibase
  10.1007/JHEP03(2014)108, 10.1007/JHEP06(2012)141} {\bibfield  {journal}
  {\bibinfo  {journal} {JHEP}\ }\textbf {\bibinfo {volume} {06}},\ \bibinfo
  {pages} {141} (\bibinfo {year} {2012})},\ \bibinfo {note} {[Addendum:
  JHEP03,108(2014)]},\ \Eprint {http://arxiv.org/abs/1205.0975}
  {arXiv:1205.0975 [hep-ex]} \BibitemShut {NoStop}%
%%CITATION = ARXIV:1205.0975;%%
\bibitem [{\citenamefont {Aaij}\ \emph {et~al.}(2016)\citenamefont {Aaij} \emph
  {et~al.}}]{Aaij:2015wpa}%
  \BibitemOpen
  \bibfield  {author} {\bibinfo {author} {\bibfnamefont {R.}~\bibnamefont
  {Aaij}} \emph {et~al.} (\bibinfo {collaboration} {LHCb}),\ }\href {\doibase
  10.1007/JHEP07(2016)052} {\bibfield  {journal} {\bibinfo  {journal} {JHEP}\
  }\textbf {\bibinfo {volume} {07}},\ \bibinfo {pages} {052} (\bibinfo {year}
  {2016})},\ \Eprint {http://arxiv.org/abs/1510.05949} {arXiv:1510.05949
  [hep-ex]} \BibitemShut {NoStop}%
%%CITATION = ARXIV:1510.05949;%%
\bibitem [{\citenamefont {Kiselev}\ \emph {et~al.}(1994)\citenamefont
  {Kiselev}, \citenamefont {Likhoded},\ and\ \citenamefont
  {Shevlyagin}}]{Kiselev:1994pu}%
  \BibitemOpen
  \bibfield  {author} {\bibinfo {author} {\bibfnamefont {V.~V.}\ \bibnamefont
  {Kiselev}}, \bibinfo {author} {\bibfnamefont {A.~K.}\ \bibnamefont
  {Likhoded}}, \ and\ \bibinfo {author} {\bibfnamefont {M.~V.}\ \bibnamefont
  {Shevlyagin}},\ }\href {\doibase 10.1016/0370-2693(94)91273-4} {\bibfield
  {journal} {\bibinfo  {journal} {Phys. Lett.}\ }\textbf {\bibinfo {volume}
  {B332}},\ \bibinfo {pages} {411} (\bibinfo {year} {1994})},\ \Eprint
  {http://arxiv.org/abs/hep-ph/9408407} {arXiv:hep-ph/9408407 [hep-ph]}
  \BibitemShut {NoStop}%
%%CITATION = HEP-PH/9408407;%%
\bibitem [{\citenamefont {Mertig}\ \emph {et~al.}(1991)\citenamefont {Mertig},
  \citenamefont {Bohm},\ and\ \citenamefont {Denner}}]{Mertig:1990an}%
  \BibitemOpen
  \bibfield  {author} {\bibinfo {author} {\bibfnamefont {R.}~\bibnamefont
  {Mertig}}, \bibinfo {author} {\bibfnamefont {M.}~\bibnamefont {Bohm}}, \ and\
  \bibinfo {author} {\bibfnamefont {A.}~\bibnamefont {Denner}},\ }\href
  {\doibase 10.1016/0010-4655(91)90130-D} {\bibfield  {journal} {\bibinfo
  {journal} {Comput. Phys. Commun.}\ }\textbf {\bibinfo {volume} {64}},\
  \bibinfo {pages} {345} (\bibinfo {year} {1991})}\BibitemShut {NoStop}%
%%CITATION = CPHCB,64,345;%%
\bibitem [{\citenamefont {Shtabovenko}\ \emph {et~al.}(2016)\citenamefont
  {Shtabovenko}, \citenamefont {Mertig},\ and\ \citenamefont
  {Orellana}}]{Shtabovenko:2016sxi}%
  \BibitemOpen
  \bibfield  {author} {\bibinfo {author} {\bibfnamefont {V.}~\bibnamefont
  {Shtabovenko}}, \bibinfo {author} {\bibfnamefont {R.}~\bibnamefont {Mertig}},
  \ and\ \bibinfo {author} {\bibfnamefont {F.}~\bibnamefont {Orellana}},\
  }\href {\doibase 10.1016/j.cpc.2016.06.008} {\bibfield  {journal} {\bibinfo
  {journal} {Comput. Phys. Commun.}\ }\textbf {\bibinfo {volume} {207}},\
  \bibinfo {pages} {432} (\bibinfo {year} {2016})},\ \Eprint
  {http://arxiv.org/abs/1601.01167} {arXiv:1601.01167 [hep-ph]} \BibitemShut
  {NoStop}%
%%CITATION = ARXIV:1601.01167;%%
\bibitem [{\citenamefont {Bodwin}\ and\ \citenamefont
  {Petrelli}(2002)}]{Bodwin:2002hg}%
  \BibitemOpen
  \bibfield  {author} {\bibinfo {author} {\bibfnamefont {G.~T.}\ \bibnamefont
  {Bodwin}}\ and\ \bibinfo {author} {\bibfnamefont {A.}~\bibnamefont
  {Petrelli}},\ }\href {\doibase 10.1103/PhysRevD.87.039902,
  10.1103/PhysRevD.66.094011} {\bibfield  {journal} {\bibinfo  {journal} {Phys.
  Rev.}\ }\textbf {\bibinfo {volume} {D66}},\ \bibinfo {pages} {094011}
  (\bibinfo {year} {2002})},\ \bibinfo {note} {[Erratum: Phys.
  Rev.D87,no.3,039902(2013)]},\ \Eprint {http://arxiv.org/abs/1301.1079}
  {arXiv:1301.1079 [hep-ph]} \BibitemShut {NoStop}%
%%CITATION = ARXIV:1301.1079;%%
\bibitem [{\citenamefont {Buckley}\ \emph {et~al.}(2015)\citenamefont
  {Buckley}, \citenamefont {Ferrando}, \citenamefont {Lloyd}, \citenamefont
  {Nordström}, \citenamefont {Page}, \citenamefont {Rüfenacht}, \citenamefont
  {Schönherr},\ and\ \citenamefont {Watt}}]{Buckley:2014ana}%
  \BibitemOpen
  \bibfield  {author} {\bibinfo {author} {\bibfnamefont {A.}~\bibnamefont
  {Buckley}}, \bibinfo {author} {\bibfnamefont {J.}~\bibnamefont {Ferrando}},
  \bibinfo {author} {\bibfnamefont {S.}~\bibnamefont {Lloyd}}, \bibinfo
  {author} {\bibfnamefont {K.}~\bibnamefont {Nordström}}, \bibinfo {author}
  {\bibfnamefont {B.}~\bibnamefont {Page}}, \bibinfo {author} {\bibfnamefont
  {M.}~\bibnamefont {Rüfenacht}}, \bibinfo {author} {\bibfnamefont
  {M.}~\bibnamefont {Schönherr}}, \ and\ \bibinfo {author} {\bibfnamefont
  {G.}~\bibnamefont {Watt}},\ }\href {\doibase 10.1140/epjc/s10052-015-3318-8}
  {\bibfield  {journal} {\bibinfo  {journal} {Eur. Phys. J.}\ }\textbf
  {\bibinfo {volume} {C75}},\ \bibinfo {pages} {132} (\bibinfo {year}
  {2015})},\ \Eprint {http://arxiv.org/abs/1412.7420} {arXiv:1412.7420
  [hep-ph]} \BibitemShut {NoStop}%
%%CITATION = ARXIV:1412.7420;%%
\bibitem [{\citenamefont {Shao}(2013)}]{Shao:2012iz}%
  \BibitemOpen
  \bibfield  {author} {\bibinfo {author} {\bibfnamefont {H.-S.}\ \bibnamefont
  {Shao}},\ }\href {\doibase 10.1016/j.cpc.2013.05.023} {\bibfield  {journal}
  {\bibinfo  {journal} {Comput. Phys. Commun.}\ }\textbf {\bibinfo {volume}
  {184}},\ \bibinfo {pages} {2562} (\bibinfo {year} {2013})},\ \Eprint
  {http://arxiv.org/abs/1212.5293} {arXiv:1212.5293 [hep-ph]} \BibitemShut
  {NoStop}%
%%CITATION = ARXIV:1212.5293;%%
\bibitem [{\citenamefont {Shao}(2016)}]{Shao:2015vga}%
  \BibitemOpen
  \bibfield  {author} {\bibinfo {author} {\bibfnamefont {H.-S.}\ \bibnamefont
  {Shao}},\ }\href {\doibase 10.1016/j.cpc.2015.09.011} {\bibfield  {journal}
  {\bibinfo  {journal} {Comput. Phys. Commun.}\ }\textbf {\bibinfo {volume}
  {198}},\ \bibinfo {pages} {238} (\bibinfo {year} {2016})},\ \Eprint
  {http://arxiv.org/abs/1507.03435} {arXiv:1507.03435 [hep-ph]} \BibitemShut
  {NoStop}%
%%CITATION = ARXIV:1507.03435;%%
\bibitem [{\citenamefont {Patrignani}\ \emph {et~al.}(2016)\citenamefont
  {Patrignani} \emph {et~al.}}]{Olive:2016xmw}%
  \BibitemOpen
  \bibfield  {author} {\bibinfo {author} {\bibfnamefont {C.}~\bibnamefont
  {Patrignani}} \emph {et~al.} (\bibinfo {collaboration} {Particle Data
  Group}),\ }\href {\doibase 10.1088/1674-1137/40/10/100001} {\bibfield
  {journal} {\bibinfo  {journal} {Chin. Phys.}\ }\textbf {\bibinfo {volume}
  {C40}},\ \bibinfo {pages} {100001} (\bibinfo {year} {2016})}\BibitemShut
  {NoStop}%
%%CITATION = CHPHD,C40,100001;%%
\bibitem [{\citenamefont {Penarrocha}\ and\ \citenamefont
  {Schilcher}(2001)}]{Penarrocha:2001ig}%
  \BibitemOpen
  \bibfield  {author} {\bibinfo {author} {\bibfnamefont {J.}~\bibnamefont
  {Penarrocha}}\ and\ \bibinfo {author} {\bibfnamefont {K.}~\bibnamefont
  {Schilcher}},\ }\href {\doibase 10.1016/S0370-2693(01)00877-2} {\bibfield
  {journal} {\bibinfo  {journal} {Phys. Lett.}\ }\textbf {\bibinfo {volume}
  {B515}},\ \bibinfo {pages} {291} (\bibinfo {year} {2001})},\ \Eprint
  {http://arxiv.org/abs/hep-ph/0105222} {arXiv:hep-ph/0105222 [hep-ph]}
  \BibitemShut {NoStop}%
%%CITATION = HEP-PH/0105222;%%
\bibitem [{\citenamefont {Aaij}\ \emph {et~al.}(2011)\citenamefont {Aaij} \emph
  {et~al.}}]{Aaij:2011jh}%
  \BibitemOpen
  \bibfield  {author} {\bibinfo {author} {\bibfnamefont {R.}~\bibnamefont
  {Aaij}} \emph {et~al.} (\bibinfo {collaboration} {LHCb}),\ }\href {\doibase
  10.1140/epjc/s10052-011-1645-y} {\bibfield  {journal} {\bibinfo  {journal}
  {Eur. Phys. J.}\ }\textbf {\bibinfo {volume} {C71}},\ \bibinfo {pages} {1645}
  (\bibinfo {year} {2011})},\ \Eprint {http://arxiv.org/abs/1103.0423}
  {arXiv:1103.0423 [hep-ex]} \BibitemShut {NoStop}%
%%CITATION = ARXIV:1103.0423;%%
\bibitem [{\citenamefont {Aaij}\ \emph {et~al.}(2015)\citenamefont {Aaij} \emph
  {et~al.}}]{Aaij:2015rla}%
  \BibitemOpen
  \bibfield  {author} {\bibinfo {author} {\bibfnamefont {R.}~\bibnamefont
  {Aaij}} \emph {et~al.} (\bibinfo {collaboration} {LHCb}),\ }\href {\doibase
  10.1007/JHEP10(2015)172} {\bibfield  {journal} {\bibinfo  {journal} {JHEP}\
  }\textbf {\bibinfo {volume} {10}},\ \bibinfo {pages} {172} (\bibinfo {year}
  {2015})},\ \Eprint {http://arxiv.org/abs/1509.00771} {arXiv:1509.00771
  [hep-ex]} \BibitemShut {NoStop}%
%%CITATION = ARXIV:1509.00771;%%
\bibitem [{\citenamefont {Berezhnoy}\ \emph {et~al.}(2011)\citenamefont
  {Berezhnoy}, \citenamefont {Likhoded}, \citenamefont {Luchinsky},\ and\
  \citenamefont {Novoselov}}]{Berezhnoy:2011xy}%
  \BibitemOpen
  \bibfield  {author} {\bibinfo {author} {\bibfnamefont {A.~V.}\ \bibnamefont
  {Berezhnoy}}, \bibinfo {author} {\bibfnamefont {A.~K.}\ \bibnamefont
  {Likhoded}}, \bibinfo {author} {\bibfnamefont {A.~V.}\ \bibnamefont
  {Luchinsky}}, \ and\ \bibinfo {author} {\bibfnamefont {A.~A.}\ \bibnamefont
  {Novoselov}},\ }\href {\doibase 10.1103/PhysRevD.84.094023} {\bibfield
  {journal} {\bibinfo  {journal} {Phys. Rev.}\ }\textbf {\bibinfo {volume}
  {D84}},\ \bibinfo {pages} {094023} (\bibinfo {year} {2011})},\ \Eprint
  {http://arxiv.org/abs/1101.5881} {arXiv:1101.5881 [hep-ph]} \BibitemShut
  {NoStop}%
%%CITATION = ARXIV:1101.5881;%%
\bibitem [{\citenamefont {Berezhnoy}\ \emph {et~al.}(2012)\citenamefont
  {Berezhnoy}, \citenamefont {Likhoded}, \citenamefont {Luchinsky},\ and\
  \citenamefont {Novoselov}}]{Berezhnoy:2012xq}%
  \BibitemOpen
  \bibfield  {author} {\bibinfo {author} {\bibfnamefont {A.~V.}\ \bibnamefont
  {Berezhnoy}}, \bibinfo {author} {\bibfnamefont {A.~K.}\ \bibnamefont
  {Likhoded}}, \bibinfo {author} {\bibfnamefont {A.~V.}\ \bibnamefont
  {Luchinsky}}, \ and\ \bibinfo {author} {\bibfnamefont {A.~A.}\ \bibnamefont
  {Novoselov}},\ }\href {\doibase 10.1103/PhysRevD.86.034017} {\bibfield
  {journal} {\bibinfo  {journal} {Phys. Rev.}\ }\textbf {\bibinfo {volume}
  {D86}},\ \bibinfo {pages} {034017} (\bibinfo {year} {2012})},\ \Eprint
  {http://arxiv.org/abs/1204.1058} {arXiv:1204.1058 [hep-ph]} \BibitemShut
  {NoStop}%
%%CITATION = ARXIV:1204.1058;%%
\bibitem [{\citenamefont {Likhoded}\ \emph {et~al.}(2016)\citenamefont
  {Likhoded}, \citenamefont {Luchinsky},\ and\ \citenamefont
  {Poslavsky}}]{Likhoded:2016zmk}%
  \BibitemOpen
  \bibfield  {author} {\bibinfo {author} {\bibfnamefont {A.~K.}\ \bibnamefont
  {Likhoded}}, \bibinfo {author} {\bibfnamefont {A.~V.}\ \bibnamefont
  {Luchinsky}}, \ and\ \bibinfo {author} {\bibfnamefont {S.~V.}\ \bibnamefont
  {Poslavsky}},\ }\href {\doibase 10.1103/PhysRevD.94.054017} {\bibfield
  {journal} {\bibinfo  {journal} {Phys. Rev.}\ }\textbf {\bibinfo {volume}
  {D94}},\ \bibinfo {pages} {054017} (\bibinfo {year} {2016})},\ \Eprint
  {http://arxiv.org/abs/1606.06767} {arXiv:1606.06767 [hep-ph]} \BibitemShut
  {NoStop}%
%%CITATION = ARXIV:1606.06767;%%
\bibitem [{\citenamefont {Lansberg}\ and\ \citenamefont
  {Shao}(2015)}]{Lansberg:2014swa}%
  \BibitemOpen
  \bibfield  {author} {\bibinfo {author} {\bibfnamefont {J.-P.}\ \bibnamefont
  {Lansberg}}\ and\ \bibinfo {author} {\bibfnamefont {H.-S.}\ \bibnamefont
  {Shao}},\ }\href {\doibase 10.1016/j.physletb.2015.10.083} {\bibfield
  {journal} {\bibinfo  {journal} {Phys. Lett.}\ }\textbf {\bibinfo {volume}
  {B751}},\ \bibinfo {pages} {479} (\bibinfo {year} {2015})},\ \Eprint
  {http://arxiv.org/abs/1410.8822} {arXiv:1410.8822 [hep-ph]} \BibitemShut
  {NoStop}%
%%CITATION = ARXIV:1410.8822;%%
\bibitem [{\citenamefont {Watanabe}\ and\ \citenamefont
  {Xiao}(2015)}]{Watanabe:2015yca}%
  \BibitemOpen
  \bibfield  {author} {\bibinfo {author} {\bibfnamefont {K.}~\bibnamefont
  {Watanabe}}\ and\ \bibinfo {author} {\bibfnamefont {B.-W.}\ \bibnamefont
  {Xiao}},\ }\href {\doibase 10.1103/PhysRevD.92.111502} {\bibfield  {journal}
  {\bibinfo  {journal} {Phys. Rev.}\ }\textbf {\bibinfo {volume} {D92}},\
  \bibinfo {pages} {111502} (\bibinfo {year} {2015})},\ \Eprint
  {http://arxiv.org/abs/1507.06564} {arXiv:1507.06564 [hep-ph]} \BibitemShut
  {NoStop}%
%%CITATION = ARXIV:1507.06564;%%
\bibitem [{\citenamefont {Lansberg}\ and\ \citenamefont
  {Shao}(2016)}]{Lansberg:2016chs}%
  \BibitemOpen
  \bibfield  {author} {\bibinfo {author} {\bibfnamefont {J.-P.}\ \bibnamefont
  {Lansberg}}\ and\ \bibinfo {author} {\bibfnamefont {H.-S.}\ \bibnamefont
  {Shao}},\ }\bibfield  {booktitle} {\emph {\bibinfo {booktitle} {{Proceedings,
  24th International Workshop on Deep-Inelastic Scattering and Related Subjects
  (DIS 2016): Hamburg, Germany, April 11-15, 2016}}},\ }\href@noop {}
  {\bibfield  {journal} {\bibinfo  {journal} {PoS}\ }\textbf {\bibinfo {volume}
  {DIS2016}},\ \bibinfo {pages} {165} (\bibinfo {year} {2016})},\ \Eprint
  {http://arxiv.org/abs/1611.02192} {arXiv:1611.02192 [hep-ph]} \BibitemShut
  {NoStop}%
%%CITATION = ARXIV:1611.02192;%%
\bibitem [{\citenamefont {Berezhnoy}\ \emph {et~al.}(1998)\citenamefont
  {Berezhnoy}, \citenamefont {Kiselev}, \citenamefont {Likhoded},\ and\
  \citenamefont {Onishchenko}}]{Berezhnoy:1998aa}%
  \BibitemOpen
  \bibfield  {author} {\bibinfo {author} {\bibfnamefont {A.~V.}\ \bibnamefont
  {Berezhnoy}}, \bibinfo {author} {\bibfnamefont {V.~V.}\ \bibnamefont
  {Kiselev}}, \bibinfo {author} {\bibfnamefont {A.~K.}\ \bibnamefont
  {Likhoded}}, \ and\ \bibinfo {author} {\bibfnamefont {A.~I.}\ \bibnamefont
  {Onishchenko}},\ }\href {\doibase 10.1103/PhysRevD.57.4385} {\bibfield
  {journal} {\bibinfo  {journal} {Phys. Rev.}\ }\textbf {\bibinfo {volume}
  {D57}},\ \bibinfo {pages} {4385} (\bibinfo {year} {1998})},\ \Eprint
  {http://arxiv.org/abs/hep-ph/9710339} {arXiv:hep-ph/9710339 [hep-ph]}
  \BibitemShut {NoStop}%
%%CITATION = HEP-PH/9710339;%%
\end{thebibliography}%

\end{document}